# Strain-Tunable Shift Current and Magneto-Optical Kerr Effect in Multiferroic Altermagnet $Fe_2Mo_3O_8$

Shengqiao Wang,[1,2] Bo Zhao,[3] Harish K. Singh,[4] Jiahao Xie,[1] Fu Li,[3,§]

Hongbin Zhang,[3] Yang Su,[2,5,†] Chen Shen,[2,*] Lijun Zhang[1]

[1]*State Key Laboratory of Integrated Optoelectronics, Key Laboratory of Automobile Materials of MOE, Key Laboratory of Material Simulation Methods & Software of MOE, School of Materials Science and Engineering, Jilin University, Changchun 130012, China*

[2]*Suzhou Laboratory, Suzhou, 215123, China*

[3]*Institute of Materials Science, Technology University of Darmstadt, 64287 Darmstadt, Germany*

[4]*Department of Chemistry and Physics of Materials,*

*University of Salzburg, Jakob-Haringer-Strasse 2a, 5020 Salzburg, Austria*

[5]*College of Physics, Jilin University, Changchun 130012, China*

[§]Contact author: fu.li@tu-darmstadt.de

[*]Contact author: shenc@szlab.ac.cn

[†]Contact author: ysu_97@163.com

**Abstract**

Altermagnetism has recently emerged as a compelling frontier in spintronics, seamlessly merging the agile tunability of ferromagnets with the hallmark merits of antiferromagnets. As a prototypical polar multiferroic featuring distinctive altermagnetism, $Fe_2Mo_3O_8$ hosts an ideal playground for exploring the intricate interplay among ferroelectric polarization, altermagnetic order, and spin-dependent responses. Here, employing first-principles calculations, we systematically investigate the coupling among polarization, spin splitting, shift current, and magneto-optical responses in $Fe_2Mo_3O_8$. Our findings reveal that switching the ferroelectric polarization not only inverts the sign of the shift current but also comprehensively reshapes the momentum-space spin-splitting texture. Furthermore, the shift current and magneto-optical spectrum exhibits strong tunability under mechanical strain. Remarkably, the application of *a*-axis uniaxial strain breaks the crystalline symmetry, thereby activating a finite magneto-optical Kerr effect that is otherwise forbidden in the pristine phase.

## Introduction

Altermagnets have recently emerged as a distinct class of magnetic materials, capturing intense interest across both spintronics and condensed matter physics [1–3]. They are characterized by compensated spin configurations with vanishing net magnetization in real space, while opposite spin sublattices are related by crystal rotations or mirror operations rather than by translation or inversion [4–6]. As a result, altermagnets break time-reversal symmetry and lift the spin degeneracy, giving rise to momentum-dependent spin splitting even in the absence of spin-orbit coupling (SOC) [7–9]. Altermagnets therefore combine key advantages of ferromagnets, such as efficient readout and controllability, with those of antiferromagnets, including the absence of stray fields and suitability for high-density integration. A variety of intriguing phenomena have already been demonstrated in altermagnetic systems, including the anomalous Hall effect [10–12], giant magnetoresistance [13], the electronic thermal Hall effect [10], chiral magnons [14], the magneto-optical Kerr effect (MOKE) [2,15], and second order nonlinear photocurrent [16–18], highlighting their broad potential for both fundamental physics and developing next-generation functional devices.

Given that the unconventional properties of altermagnets are intimately governed by crystal symmetry, external perturbations capable of modifying the lattice environment provide an effective route for tailoring their electronic, magnetic, and transport properties [19–21]. Among the various external stimuli, strain engineering has emerged as one of the most effective and experimentally accessible approaches for tuning the electronic and magnetic properties of altermagnets. [22,23]. Recent studies have demonstrated that strain provides a versatile means of manipulating both crystal symmetry and electronic structure, thereby enabling the control of a wide range of altermagnetic phenomena. In CrSb and $RuO_2$, strain-induced symmetry reduction and Berry curvature redistribution significantly modulate the allowable anomalous Hall response [24,25]. Furthermore, mechanical strain can fundamentally alter the altermagnetic spin-splitting texture, as exemplified by the shear-strain-driven transition from *g*-wave to *d*-wave altermagnet in CrSb [26], as well as lifting symmetry-protected valley degeneracy to induce valley polarization in two dimensional altermagnets [23,27]. Moreover, by breaking additional symmetries, strain can activate otherwise forbidden magneto-optical responses, including the MOKE [25]. These findings establish strain engineering as an effective strategy for tailoring the transport, magnetic, optical, and electronic properties of altermagnetic system.

In this context, $Fe_2Mo_3O_8$ provides a representative platform for investigating the coupled control of altermagnetic and optical responses in multiferroic materials. This system uniquely integrates intrinsic ferroelectric polarization [28,29], magnetoelectric-coupled multiferroicity [30], pronounced optical response [31], and potentially accessible polarization-switching pathway [32]. Because both the nonrelativistic spin splitting and the optical response tensors are highly sensitive to symmetry, $Fe_2Mo_3O_8$ is well suited for revealing how polarization reversal and strain can cooperatively modulate altermagnetic spin splitting textures and optical responses. Here, employing first-principles calculations, we investigate the interplay among polarization, spin splitting, shift current, and MOKE in polar altermagnet $Fe_2Mo_3O_8$. We find that $Fe_2Mo_3O_8$ exhibits *g*-wave altermagnetic spin splitting texture, while polarization reversal reconstructs the spin splitting texture and reverses the shift current polarity. Additionally, strain strongly modifies the shift current spectrum, and *a*-axis uniaxial strain activates a finite MOKE absent in the pristine structure. These results establish $Fe_2Mo_3O_8$ as a promising platform for coupling polarization, second order nonlinear photocurrent, and strain-tunable magneto-optical responses.

**Computational Methods**

First-principles density functional theory (DFT) calculations were carried out using the Vienna *ab initio* Simulation Package (VASP) [33–37]. The electron-ion interactions were described by the projector augmented-wave (PAW) method, and the exchange-correlation functional was treated within the generalized gradient approximation (GGA) using the Perdew-Burke-Ernerhof (PBE) parametrization [38]. To accurately capture the strong electronic correlations associated with the localized Fe *d* states, the Dudarev formulation of the DFT+*U* method was employed, incorporating an effective Hubbard parameter $U_{\text{eff}} = 4$ eV for the Fe *d* orbitals [39]. The plane-wave basis cutoff energy was set to 500 eV. The convergence criteria for the electronic self-consistent loop and the relaxation forces were strictly set to $10^{-6}$ eV and 0.01 eV/Å, respectively. Brillouin zone integration was performed utilizing a Γ-centered 5×5×3 Monkhorst–Pack *k*-point mesh in relaxation and 10×10×6 in self-consistent loop.

The shift current was calculated via maximally localized Wannier functions constructed using the Wannier90 code package [40], subsequent interpolation and integration were performed over a dense 100×100×100 *k*-point mesh. The frequency-dependent shift current conductivity tensor was evaluated within the independent-particle approximation following the Sipe-Shkrebtii formalism, using a Wannier interpolation implementation [41,42]. The dc photocurrent induced by monochromatic light is written as

$$j^a = 2\text{Re}[\sigma^{abc}(0;\omega,-\omega)E^b(\omega)E^c(-\omega)]$$

where *a*, *b*, and *c* denote Cartesian directions. The interband contribution to the shift current is given by

$$\sigma^{abc}(0;\omega,-\omega) = -\frac{i\pi e^3}{4\hbar^2}\int_{\text{BZ}}\frac{d^3k}{(2\pi)^3}\sum_{n,m} f_{nm}\left(I_{mn}^{abc} + I_{mn}^{acb}\right)[\delta(\omega_{mn}-\omega) + \delta(\omega_{nm}-\omega)],$$

with

$$I_{mn}^{abc} = r_{mn}^{b} r_{nm}^{c;a}$$

Here $r_{nm}^{a} = (1 - \delta_{nm})A_{nm}^{a}$, $A_{nm}^{a} = i\langle u_{nk} | \partial_{k_a} u_{mk}$, and the generalized derivative is defined as

$$r_{nm}^{a;b} = \partial_{k_b} r_{nm}^{a} - i(A_{nn}^{b} - A_{mm}^{b}) r_{nm}^{a}$$

The polar MOKE spectrum was subsequently derived from the diagonal and off-diagonal components of this tensor. The dielectric tensor and the corresponding MOKE response were derived utilizing a 10×10×6 *k*-point mesh. The Kerr rotation angle and ellipticity are given by the following expressions [43–45]:

$$\theta_K(\omega) + i\eta_K(\omega) = -\frac{\varepsilon_{xy}(\omega)}{[\varepsilon_{xx}(\omega) - 1]\sqrt{\varepsilon_{xx}(\omega)}}$$

Here, $\varepsilon_{xy}$ and $\varepsilon_{xx}$ denote the off-diagonal and diagonal components of the complex dielectric tensor, respectively. In this work, we focus on the polar Kerr geometry, in which the magnetization is oriented perpendicular to the reflecting surface of the sample. By comparing the strain-dependent evolution of $\varepsilon_{xy}$ and the corresponding $\theta_K$ spectra, we analyze the strain-induced activation of the MOKE from zero to finite values.

**Results**

Belonging to the $A_2Mo_3O_8$ family, $Fe_2Mo_3O_8$ crystallizes in the polar hexagonal space group $P6_3mc$. The $Fe^{2+}$ ions occupy both tetrahedral and octahedral oxygen coordination sites, which form an alternating honeycomb arrangement when projected onto the *ab* plane, as shown in Fig. 1(a) and 1(b). These layers are separated along the *c*-axis by nonmagnetic $Mo^{4+}$ ions arranged in $Mo_3O_8$ trimers, giving rise to an intrinsic structural polarization along the *c* direction [28], with a reported value as high as 230 $\mu C/cm^2$ [29]. Experimentally, $Fe_2Mo_3O_8$ undergoes an antiferromagnetic transition at a Néel temperature ($T_N$) of approximately 60 K [46]. The system also exhibits pronounced magnetoelectric effects [47], and hosts multiple magnetic states, with a magnetic structure that is tunable by external fields or chemical doping [28,48]. Previous first-principles calculation further showed that the altermagnetic state (ALM) is energetically favored over the ferromagnetic (FM) and compensated ferrimagnetic state (CFIM) in the pristine structure [32]. We therefore take the ALM configuration as the magnetic reference state for the following research. Fig. 1(c) further shows the ALM configuration of $Fe_2Mo_3O_8$ features incompletely compensated antiparallel magnetic moments between tetrahedrally and octahedrally coordinated Fe atoms within each layer, while the finite layer moments are cancelled by opposite moments on the corresponding Fe sites in adjacent layers, yielding zero net magnetization macroscopically. $Fe_2Mo_3O_8$ satisfies the symmetry criteria for an altermagnet: the opposite-spin sublattices are not related by an inversion $[C_2 \| \mathcal{P}]$ or by a pure translation $[C_2 \| \tau]$. Here, $C_2$ denotes a 180° spin rotation, $\mathcal{P}$ denotes spatial inversion, and $\tau$ denotes a spatial translation. Instead, the opposite-spin lattices are connected by rotational and mirror symmetry operations. A representative spin-space symmetry operation is [$C_2$ || $C_{2_001}$ | 0 0 ½], as shown in Fig. 1(c).

To identify the altermagnetic signatures of $Fe_2Mo_3O_8$, we performed a high-symmetry-preserving structural relaxation on the pristine phase, and further calculated the spin-polarized eigenvalues along the boundaries of the three Brillouin-zone planes are shown in Fig. 1(d). Consistent with previous reports, the band structure of $Fe_2Mo_3O_8$ exhibits an indirect band gap of approximately 1.1 eV, with the valence-band maximum at K and the conduction-band minimum at M, while remains degenerate on the nodal plane (Γ-M-K-Γ) and the Brillouin-zone boundaries (A-L-H-A) due to the protection of the corresponding little-group symmetries [30]. In contrast, a clear nonrelativistic spin splitting emerges along the interior D-U-P-D path on the $k_z$=0.25 plane, providing a clear fingerprint of altermagnetism, as shown in Fig. 1(e). After SOC-inclusive relaxation, the macroscopic net magnetization remains zero, whereas the lowered effective symmetry induces additional spin splitting along the originally degenerate high-symmetry paths (see the Figure S1).

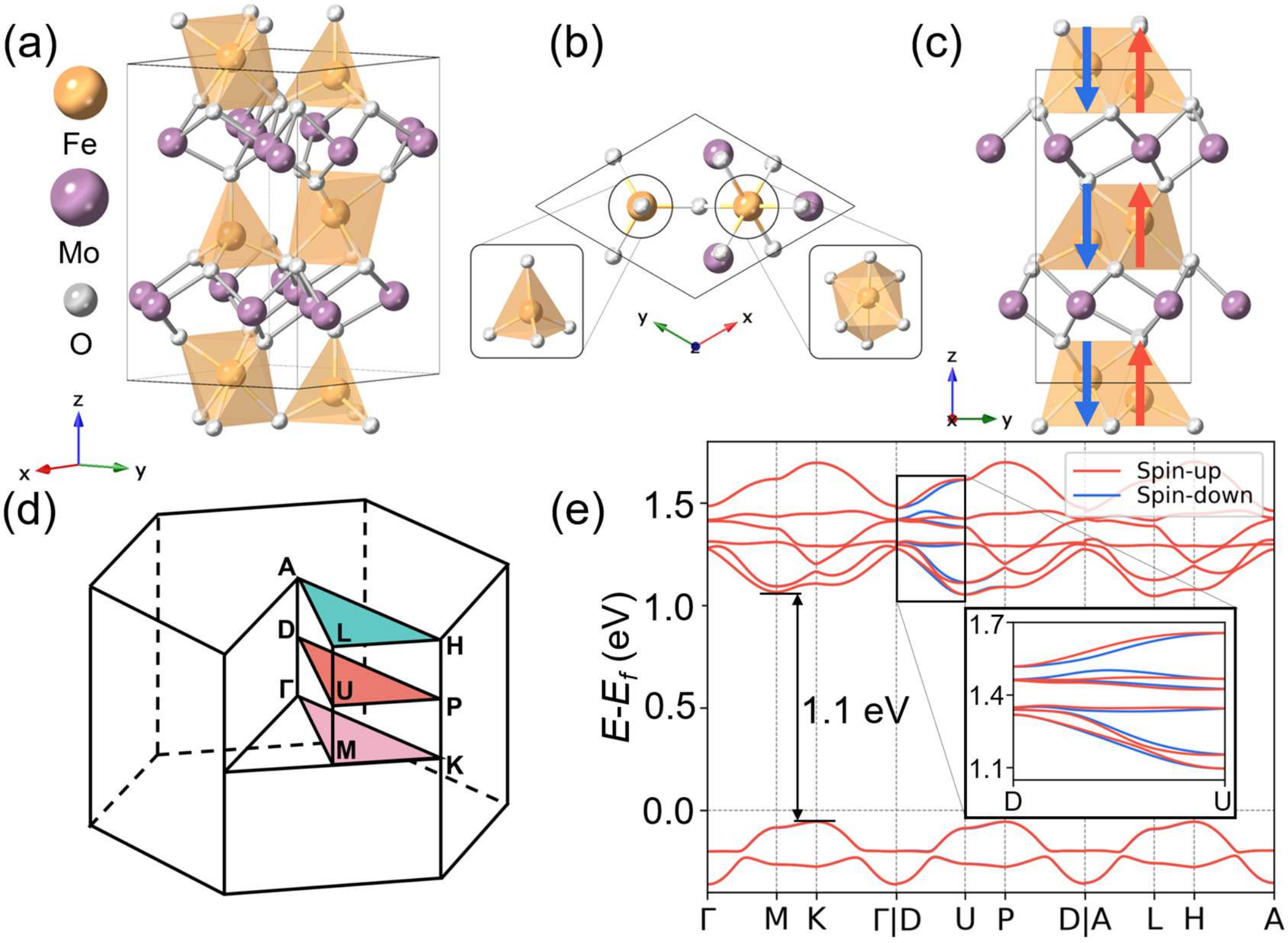

Figure 1. Crystal structure, compensated magnetic order, and altermagnetic electronic structure of $Fe_2Mo_3O_8$. (a) Schematic crystal structure of $Fe_2Mo_3O_8$, showing the spatial stacking of Fe–O coordination polyhedra and $Mo_3O_8$ layered units along the polar axis. (b) Projection of the Fe sublattice onto the *ab* plane, highlighting the alternating arrangement of tetrahedral $Fe_T$ and octahedral $Fe_O$ sites. (c) AFM magnetic configuration of $Fe_2Mo_3O_8$, in which adjacent magnetic

layers carry opposite spin orientations, resulting in zero macroscopic net magnetization. (d) the Brillouin zone of $Fe_2Mo_3O_8$ used for the band structure analysis with the high-symmetry points indicated. (e) Spin polarized band structures of the high-symmetry reference structure.

Previous studies identified a feasible polarization-switching pathway in $Fe_2Mo_3O_8$, providing the structural basis for electrically driven polarization reversal [32]. In the pristine $Fe_2Mo_3O_8$, the spontaneous polarization is oriented along the *c*-axis, consistent with the polar 6mm point group symmetry, under which the *c*-axis polar-vector component remains invariant. Motivated by recent proposals of polarization-controlled altermagnets [30,49,50], we investigate how polarization switching modulates both the altermagnetic spin splitting and the nonlinear optical response in $Fe_2Mo_3O_8$. The left panels of Figs. 2(a) and 2(b) show the configurations of polarization up and down, respectively, where polarization switching exchanges the relative positions of octahedrally and tetrahedrally coordinated Fe atoms along the polar axis, while the weak ferroelectric-magnetic coupling in type-I multiferroic $Fe_2Mo_3O_8$ preserves the local Fe magnetic moment directions [51,52], leading to an interchange of the spin-up and spin-down eigenvalue channels in the right panels while maintaining zero macroscopic net magnetization. This behavior is further manifested in the spin splitting maps on the $k_z$=0.25 plane where the alternating positive and negative petal-like regions form a *g*-wave pattern consistent with the $C_3$ symmetry and are reversed upon polarization switching as shown in Fig. 2(c). A corresponding reversal is also observed in the nonlinear optical response. As shown in Fig. 2(d), the dominant shift current tensor component $\sigma_{zyy}$, changes sign upon polarization reversal while largely preserving its spectral shape and magnitude. These results demonstrate that the shift current provides a direct optical fingerprint of polarization-controlled reversal of the altermagnetic spin splitting in in $Fe_2Mo_3O_8$.

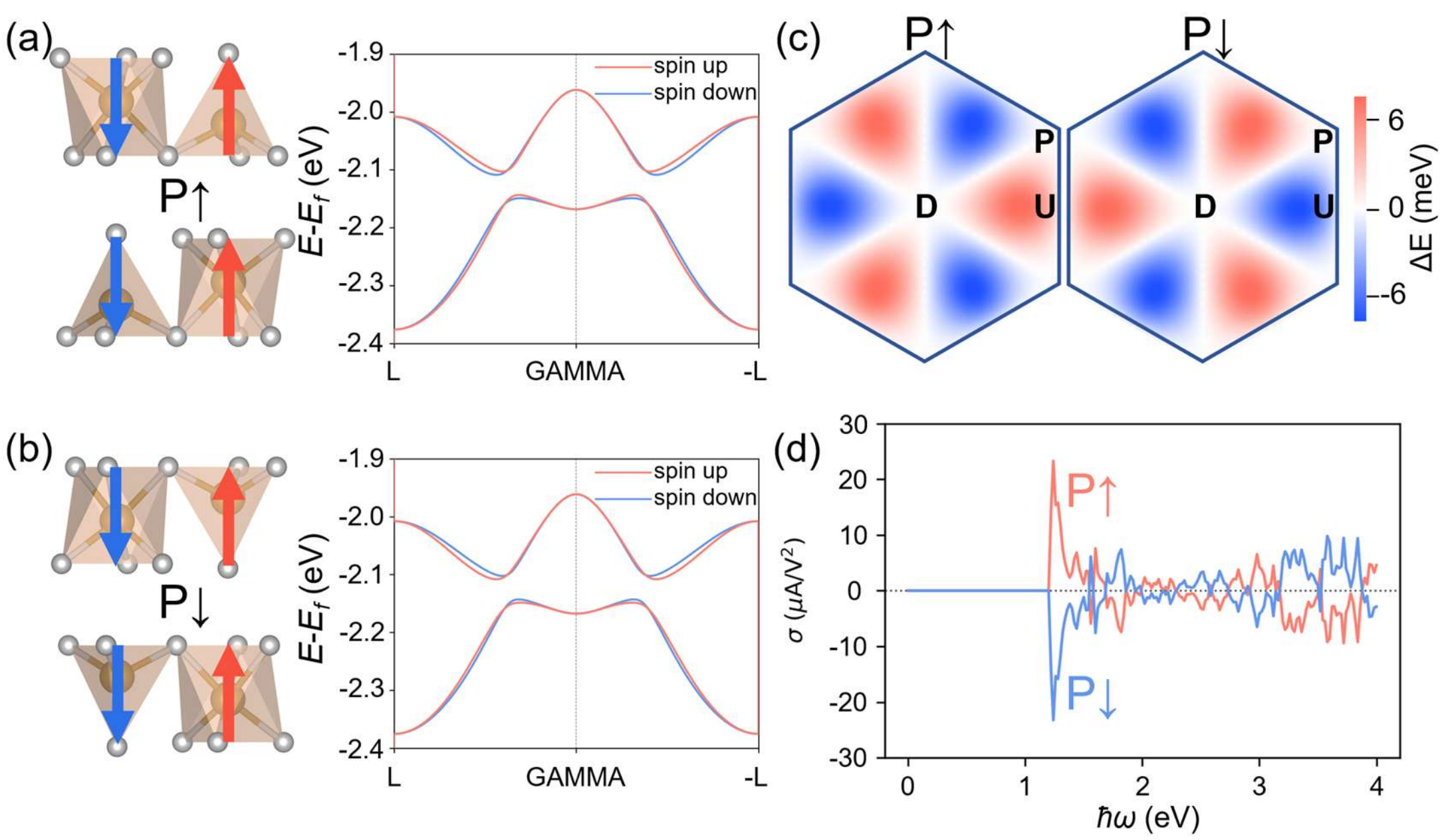

Figure 2. Polarization-controlled reversal of the nonrelativistic spin splitting and shift current in altermagnet $Fe_2Mo_3O_8$. (a, b) Crystal and spin configurations for the two opposite ferroelectric polarization states, $P\uparrow$ and $P\downarrow$, together with the corresponding spin-polarized band structures. Upon polarization switching, the spin-up and spin-down channels are interchanged, indicating a complete reversal of the $k$-space nonrelativistic spin splitting. (c) Calculated distribution of the spin splitting observed in the VBM on the $k_z$=0.25 plane of the Brillouin zone for the $P\uparrow$ and $P\downarrow$ states. The pattern is reversed after polarization flipping. (d) The shift current spectra of component $\sigma_{zyy}$ for the two opposite ferroelectric polarization states, $P\uparrow$ and $P\downarrow$.

In intrinsically non-centrosymmetric magnetic materials, subtle changes in lattice parameters alter not only the bandgap and band-edge dispersion, but also orbital hybridization, interband transition channels, and the associated geometric phase properties, thereby strongly affecting the optical responses [53–55]. Therefore, to further explore the tunability of the optical responses in $Fe_2Mo_3O_8$, three representative strain modes were considered, $a$-axis uniaxial strain, in-plane biaxial strain (denoted as $ab$ strain), and shape-preserving strain applied along the $a$, $b$, and $c$ directions (denoted as $abc$ strain), each with theoretical strain amplitudes ranging from -5% to +5%. We then tracked their effects on crystal symmetry, allowed optical tensor components, shift current magnitudes, and MOKE activation. Rather than producing a uniform response, these strain modes fall into two distinct categories. The first, represented mainly by $ab$ strain and $abc$ strain, primarily reshapes the magnitude and spectral distribution of symmetry-allowed shift current channels while leaving the off-diagonal optical response largely suppressed. The second, represented by $a$-axis uniaxial strain, lowers the symmetry of the pristine phase strongly, lifts the constraints on the off-diagonal optical tensor, and thereby activates a finite MOKE response. Table 1 summarizes representative configurations selected from the full strain screening, with the complete results provided in Tables S2 and S3. The +5% $a$-axis strain lowers the magnetic point group to 2′ and activates a sizable Kerr rotation, while the +4% $ab$ strain preserves the pristine point group 6mm and enhances shift current, and although +2% $abc$ strain is imposed as a symmetry-preserving mode, the fully relativistic relaxation leads to a slight SOC-induced symmetry lowering with a detected magnetic point group 2′, while still yielding the largest shift current among the screened structures.

| **Strain Mode** | **Ratio (%)** | **Point Group** | **MSG/MPG** | $\sigma_{zyy}$ **Max ($\mu$A/V$^2$)** | $\theta_K$ **Max (deg)** |
|---|---|---|---|---|---|
| pristine | 0 | 6mm | $P6_3$′m′c/6′mm′ | 23.4 | / |
| *a* | +5 | 2 | $P2_1$′/2′ | 20.0 | 3.13 |
| *ab* | +4 | 6mm | $P6_3$′m′c/6′mm′ | 44 | / |
| *abc* | +2 | 2 | $P2_1$′/2′ | 52.1 | -0.85 |

Table 1. Summary of representative strained $Fe_2Mo_3O_8$ structures and their calculated optical responses. The crystallographic point groups and the magnetic space group (MSG) / point group (MPG) are provided to facilitate the symmetry-related discussion of the strain-modulated shift current and MOKE responses. The complete strain-dependent results, including all ratios and allowed shift current tensor components, are provided in Tables S2 and S3.

To further elucidate the microscopic origin of the shift current under the influence of strain, we applied the symmetry-preserving *ab* and *abc* strain modes shown in Fig. 3(a). Following full structural relaxation, the *ab* +4% and *abc* +2% configurations were selected as representative cases that most effectively maximize the $\sigma_{zyy}$ shift current response; their calculated spectra are contrasted with the pristine phase in Fig. 3(b). In the pristine structure, the dominant shift current response appears near the absorption onset, with a peak around $\hbar\omega \approx 1.24$ eV and reaching a magnitude of about 23 $\mu A/V^2$. Under *ab* +4% strain, the main response is transferred to the higher-energy region, where the peak near $\hbar\omega \approx 3.00$ eV reaches approximately 44 $\mu A/V^2$. An even larger response is obtained under +2% *abc* strain, for which the peak magnitude reaches about 52 $\mu A/V^2$ around $\hbar\omega \approx 3.12$ eV. These results show that strain does not simply scale the original low-energy response, but substantially reshapes the shift current spectrum by creating dominant high-energy photocurrent peaks. To clarify the origin of the enhanced shift current, we compare the joint density of states (JDOS) in Fig. 3(c) and *k*-resolved contributions in Figs. 3(d)-(f). For pristine structure, the JDOS peak peak appears near 1.24 eV, coinciding with the dominant $\sigma_{zyy}$ shift current peak; similarly, for the +4% *ab* and +2% *abc* structures, the JDOS peak at 3.03 and 3.12 eV closely coincide with the corresponding shift current features, indicating that the strain-enhanced shift current is associated with an increased number of available interband channels transition at the relevant photon energies. The associated *k*-resolved $\sigma_{zyy}$ distributions, obtained by summing over $k_z$ and projecting onto the $k_z$=0 plane, reveal how these transitions contribute to the net response. In pristine structure, the *k*-resolved shift current contribution is mainly confined to a localized region near the K point, with relatively weak intensity. By contrast, the +4% *ab* strain produces broadly distributed *k*-space contributions throughout the Brillouin zone, while the +2% *abc* strain shifts the dominant contribution from K to the vicinity of Γ. These distinct redistribution patterns, together with the increased number of available interband transition channels, demonstrate that global lattice modulation can strongly enhance the shift current while simultaneously reshaping its spectral and momentum-space distributions, thereby providing multiple routes for tuning the second-order nonlinear optical response of altermagnet $Fe_2Mo_3O_8$.

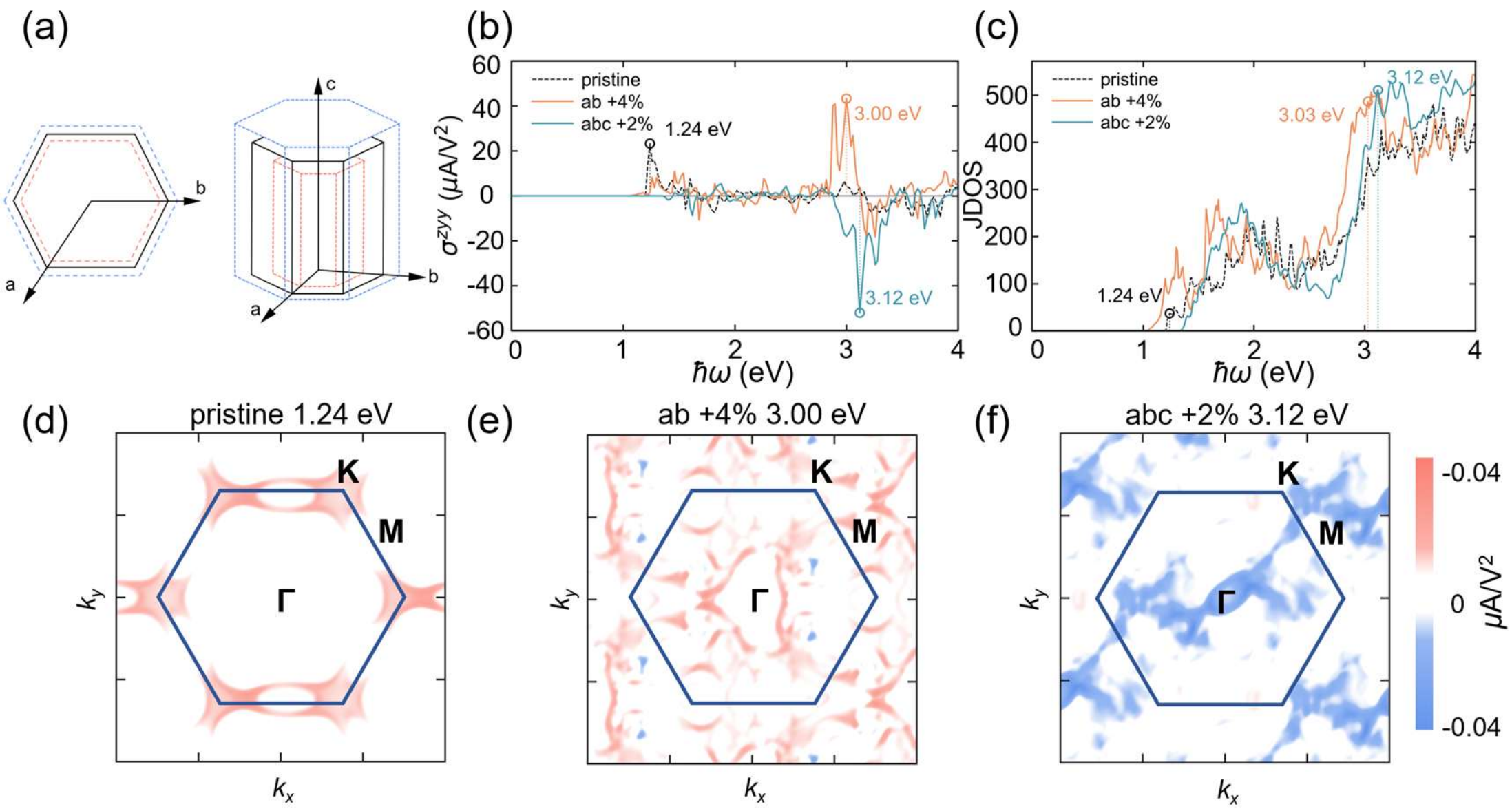

Figure 3. Strain engineering of shift current and *k*-resolved contributions in $Fe_2Mo_3O_8$. (a) Schematic illustration of the two representative strain modes used to enhance the shift current response: in-plane *ab* biaxial strain (denoted as *ab* strain) and shape-preserving strain applied simultaneously along the *a*, *b*, *c* directions (denoted as *abc* strain). (b) Calculated $\sigma_{zyy}$ shift current spectra for pristine $Fe_2Mo_3O_8$, +4% *ab* strain, and +2% *abc* strain. The two strained configurations correspond to the largest shift current enhancements among the screened strain conditions, with the dominant responses shifted to the high-energy region. (c) JDOS spectra for the same three structures. The enhanced JDOS around the photon energies corresponding to the main shift current peaks indicates that strain provides a high density of available interband transition channels, which is a necessary condition for the enlarged nonlinear photocurrent response. (d-f) *k*-resolved $\sigma_{zyy}$ shift current contributions for pristine $Fe_2Mo_3O_8$, +4% *ab* strain and +2% *abc* strain, evaluated at the representative peak energies of 1.24, 3.00, and 3.12 eV, respectively. The *k*-resolved contributions are summed over $k_z$ from 0 to 1 and projected onto the $k_z$=0 plane, with the integrated values consistent with the corresponding total shift current amplitudes.

Beyond spectral tuning, strain-induced symmetry lowering can qualitatively modify the optical response by changing the allowed tensor components. The MOKE response is governed by the antisymmetric off-diagonal dielectric component $\varepsilon_{xy}$, where $\varepsilon_{xy} = -\varepsilon_{yx}$. For the pristine $Fe_2Mo_3O_8$, the magnetic point group is 6′mm′. Symmetry operations such as $2'_{001}$ transform this component as $\varepsilon_{xy} \rightarrow -\varepsilon_{xy}$; therefore, invariance under the magnetic point group requires $\varepsilon_{xy} = 0$, making the MOKE response symmetry-forbidden. While as schematically shown in Fig. 4(a), uniaxial strain ranging from -5% to +5% was schematically applied along the *a*-axis to deliberately lower the crystal symmetry of $Fe_2Mo_3O_8$, reducing the symmetry from 6′mm′ to 2′, thereby removing the symmetry constraint on $\varepsilon_{xy}$ and allowing a finite MOKE.

Figs. 4(b) and 4(c) show the MOKE rotation angle spectra assuming incident light along the polar *c*-axis under progressively increasing tensile and compressive *a*-axis uniaxial strains, respectively. Consistent with the symmetry analysis, an *a*-axis strain as small as 1% already produces a finite MOKE signal, while the peak Kerr rotation increases with the strain ratio over the examined range and reaches nearly 3° under +5% tensile strain. Although finite MOKE signals are listed for several nominally symmetry-preserving strained structures, these residual values originate from relaxation-induced symmetry lowering rather than from the symmetry character of the imposed strain mode itself, and are therefore not considered robust MOKE activation channels. Across all strain ratios considered here, the system retains its altermagnetic signature (see the Figures S6-S8), demonstrating that symmetry-breaking strain provides an effective route for tuning the magneto-optical response of multiferroic altermagnet $Fe_2Mo_3O_8$.

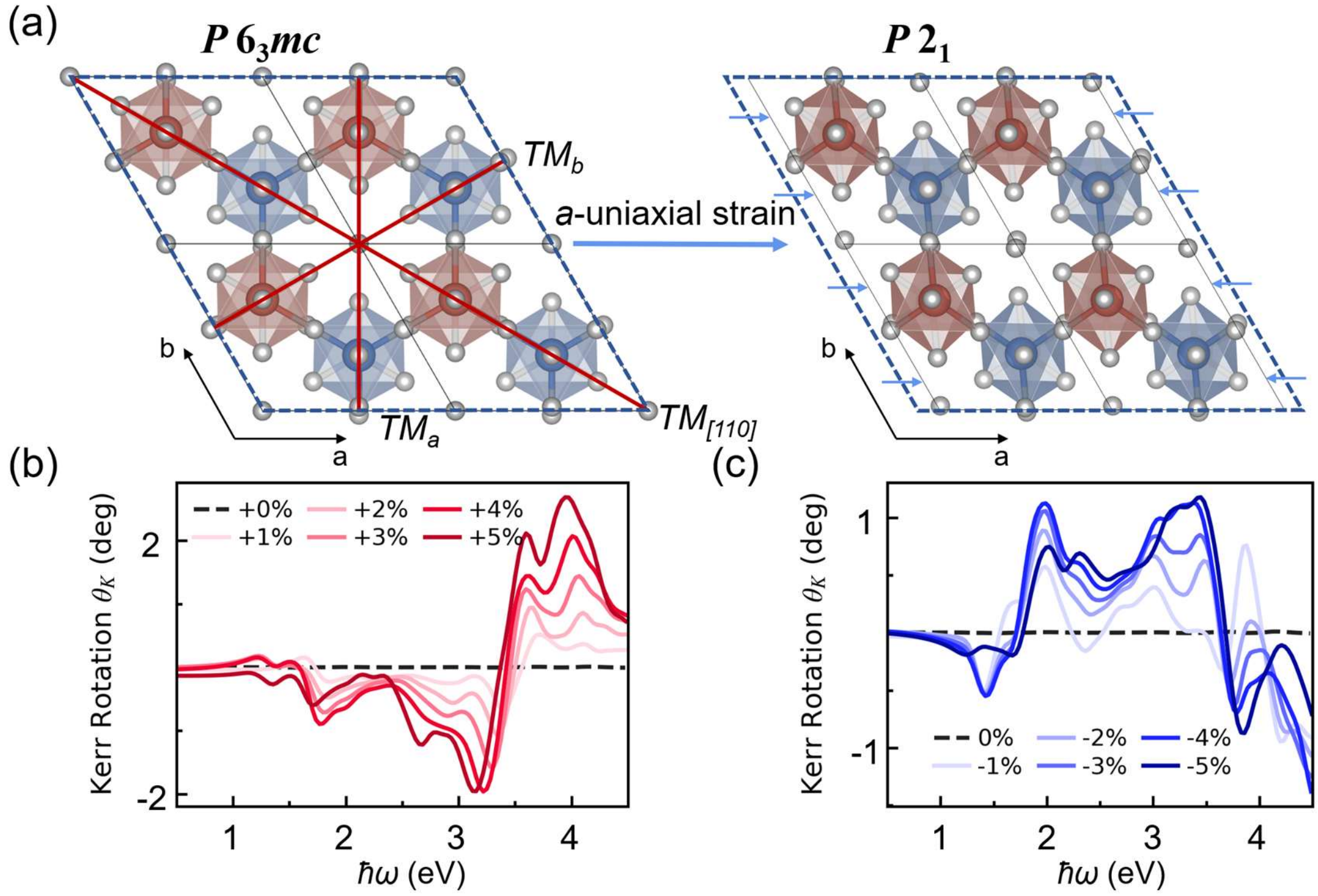

Figure 4. Symmetry-breaking activation of MOKE by *a*-axis uniaxial strain in $Fe_2Mo_3O_8$. (a) Schematic illustration of the symmetry change induced by *a*-axis uniaxial strain. In the pristine structure, the relevant mirror-related constraints suppress the off-diagonal optical response, leading to a negligibly small MOKE signal. Applying *a*-axis uniaxial strain explicitly removes these constraints and enables a finite polar MOKE response. The red and blue Fe-centered polyhedra denote Fe sites with magnetic moments pointing upward and downward, respectively. (b, c) Kerr rotation spectra calculated for tensile and compressive *a*-axis uniaxial strains, respectively, with light incident along the polar *c*-axis. The dashed black line represents the pristine structure, for which the MOKE signal remains negligible. Both signs of *a*-axis strain generate a finite Kerr rotation, demonstrating controlled symmetry-breaking activation of the MOKE response.

## Conclusion

Using first-principles calculations, we elucidate the connections among ferroelectric polarization, altermagnetic spin splitting texture, and optoelectronic response in the multiferroic altermagnetic $Fe_2Mo_3O_8$. In the high-symmetry reference structure, $Fe_2Mo_3O_8$ exhibits a *g*-wave altermagnetic spin splitting pattern. We show that ferroelectric polarization reversal reconstructs the spin-splitting texture while preserving the compensated net magnetic order, and simultaneously reverses sign of the shift current. These results reveal a clear coupling among polarization, spin splitting, and nonlinear photocurrent in this system. Furthermore, strain strongly modulates the optoelectronic response of $Fe_2Mo_3O_8$. Biaxial and shape-preserving strains enhance the shift current and reshape its spectral distribution, with the peak shift current responses reaching 44.0 and 52.1 $\mu A/V^2$ under +4% *ab* and +2% *abc* strains, respectively, whereas +5% tensile *a*-axis uniaxial strain lowers the symmetry and activates a finite MOKE response with a Kerr rotation of up to 3.13°, which is absent in the pristine state. Taken together, these results identify $Fe_2Mo_3O_8$ as a promising multiferroic altermagnetic platform in which polarization, spin-dependent electronic structure, and optical responses can be coupled and tuned, highlighting its potential for multifunctional optoelectronic and spintronic applications.

## Acknowledgement

The authors thank the NHR Center NHR4CES at TU Darmstadt and the High-Performance Computing Center of Jilin University for allocating computational resources on the high-performance computer Lichtenberg and local HPC facilities, respectively. This work was supported by the National Natural Science Foundation of China under Grant Nos. 62321166653 and 62125402.